\documentclass[11pt]{article}
\usepackage{indentfirst}
\usepackage{float}

\usepackage{tikz}
\usetikzlibrary{shapes, fit, backgrounds, calc, positioning}
\usetikzlibrary{arrows.meta}
\usetikzlibrary{decorations.pathmorphing}

\usepackage[utf8]{inputenc}
\usepackage{amsmath}
\usepackage{amssymb}
\usepackage{amsthm}
\usepackage{cite}
\usepackage{mciteplus}
\usepackage{braket}
\usepackage{booktabs}
\usepackage{xcolor}

\newtheorem{theorem}{Theorem}
\newtheorem{corollary}{Corollary}
\newtheorem{definition}{Definition}
\newtheorem{lemma}{Lemma}

\newtheorem{example}{Example}
\newtheorem{remark}{Remark}

\newcommand\blfootnote[1]{%
  \begingroup
  \renewcommand\thefootnote{}\footnote{#1}%
  \addtocounter{footnote}{-1}%
  \endgroup
}

\newcommand{\F}{\mathbb{F}}

\newcommand{\cE}{\mathcal{E}}

\newcommand{\cP}{\mathcal{P}}

\newcommand{\cG}{\mathcal{G}}

\newcommand{\GHZ}{\mathrm{GHZ}}
\newcommand{\CZ}{C^{\mathrm Z}}
\newcommand{\ED}{E_{\mathrm D}}

\DeclareMathOperator{\Tr}{Tr}

\title{The Capacity of Conference Key Agreement\\ over GHZ Networks}
\author{Hua Sun}
\date{}

\begin{document}
\maketitle

\blfootnote{
Hua Sun (email: hua.sun@unt.edu) is with the Department of Electrical Engineering at the University of North Texas.}

\begin{abstract}
$K$ users share independent GHZ states, each held by a subset of the users, and want a common secret key, using local operations and public classical communication (LOCC). We show that the conference key capacity, the largest number of secret key bits per copy of the resource, equals the key capacity of the classical source obtained by measuring every GHZ state. Measuring first is therefore optimal, even when the users keep quantum memory and communicate over any number of rounds. The capacity also equals the GHZ distillation capacity, the largest number of GHZ states shared by all users obtainable per copy.
\end{abstract}

\allowdisplaybreaks
\emergencystretch=3em
\section{Introduction}\label{sec:intro}

A Bell pair gives two users one common secret bit, and a GHZ state gives any number of users one. The problem studied here starts when each available GHZ state is held by only some of the users, while all $K$ users want the same key, a \emph{conference key}. The users are labeled $1,\dots,K$, and the resource is described by a \emph{hypergraph} $\cG=([K],\cE)$, a list $\cE$ of \emph{edges}, each edge $e$ a subset of at least two users. The users hold independent GHZ states, one for each edge. The GHZ state on edge $e$ gives one qubit to each user in $e$, and measuring these qubits gives all of them the same uniform random bit, the \emph{label} of $e$; the users in $e$ are its \emph{holders}. The users may apply any local quantum operations, keep quantum memory, and broadcast classical messages over an authenticated public channel, over any number of rounds, each message depending on all earlier ones; an eavesdropper observes all of this public communication. The \emph{conference key capacity} $C$ is the largest number of common secret bits per copy of the resource, and the \emph{GHZ distillation capacity} $\ED$ is the largest number of global GHZ states, shared by all $K$ users, per copy. Quantum conference key agreement is reviewed in \cite{Murta20}.

\begin{figure}[H]
\centering
\begin{tikzpicture}[scale=1, every node/.style={font=\small},
  user/.style={circle, draw, thick, minimum size=7mm, inner sep=0pt},
  bell/.style={thick, decorate, decoration={snake, amplitude=0.6mm, segment length=3mm}},
  ann/.style={rectangle, rounded corners=2pt, draw, fill=black!4, inner sep=2pt, font=\footnotesize}]
\begin{scope}
\node[user] (a) at (0,1.9) {$1$};
\node[user] (b) at (-1.6,0) {$2$};
\node[user] (c) at (1.6,0) {$3$};
\draw[bell] (a) -- node[left=3pt] {$\{1,2\}$} (b);
\draw[bell] (a) -- node[right=3pt] {$\{1,3\}$} (c);
\draw[bell] (b) -- node[below=3pt] {$\{2,3\}$} (c);
\node[align=center] at (0,-1.15) {(a) any LOCC with quantum memory\\ and public communication; one common secret key};
\end{scope}
\begin{scope}[xshift=7.4cm]
\node[user] (a) at (0,1.9) {$1$};
\node[user] (b) at (-1.6,0) {$2$};
\node[user] (c) at (1.6,0) {$3$};
\draw[thick] (a) -- node[left=2pt] {$a_1,a_2$} (b);
\draw[thick] (a) -- node[right=2pt] {$b_1,b_2$} (c);
\draw[thick] (b) -- node[below=2pt] {$c_1,c_2$} (c);
\node[ann, above=1mm of a] {$a_1+b_1$};
\node[ann, left=1mm of b] {$a_2+c_1$};
\node[ann, right=1mm of c] {$b_2+c_2$};
\node[align=center] at (0,-1.15) {(b) measure, broadcast one parity each;\\ key $(a_1,a_2,b_2)$, rate $3/2$};
\end{scope}
\end{tikzpicture}
\caption{The Bell triangle. (a) The resource and the requirement; every cut crosses two pairs, so the minimum cut is $2$. (b) Measuring each pair and broadcasting one parity per user over two copies achieves $3/2$, which Theorem~\ref{thm:main} shows is the capacity.}
\label{fig:triangle}
\end{figure}
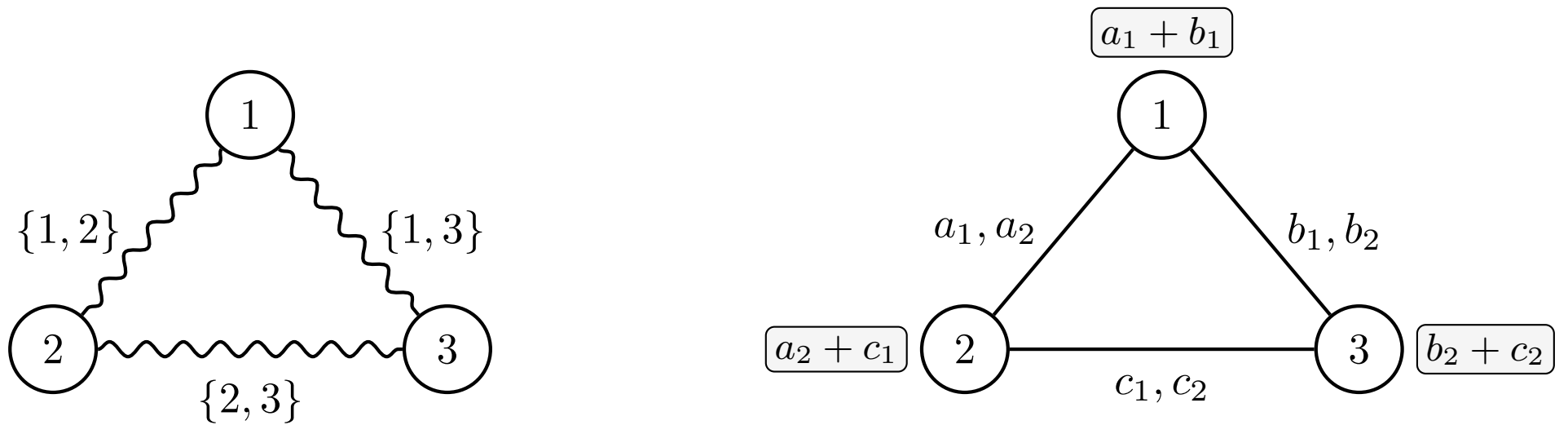

The smallest instance already shows the difficulty. In the \emph{Bell triangle} (Figure~\ref{fig:triangle}(a)), users $1,2,3$ share one Bell pair on each pair of users. Measuring every pair gives uniform bits $a$, $b$ and $c$, known to the pairs $\{1,2\}$, $\{1,3\}$ and $\{2,3\}$, respectively. Over two copies, one parity broadcast by each user lets every user recover all six bits and leaves three secret bits (Figure~\ref{fig:triangle}(b)), rate $3/2$. A \emph{cut} divides the users into two nonempty groups. A key common to all users is in particular a key between the two groups, so the capacity is at most the entanglement across the smallest cut, the \emph{minimum cut}, which is $2$ here. Vrana and Christandl showed that stochastic LOCC extracts GHZ states from any cycle of Bell pairs at rate $2$, with a success probability that may vanish, and asked whether LOCC with vanishing error also achieves rate $2$ \cite{VranaChristandl17}. Every converse for unrestricted LOCC that we know of, from relative entropy \cite{Das21, Neumann26}, squashed entanglement \cite{BaumlAzuma17} and Partovi's monotone \cite{Partovi09} for $C$, and from multipartite Rains measures \cite{Murray26} for $\ED$, which bounds $C$ through Lemma~\ref{prop:equal}, gives an upper bound that is at least the minimum cut on such networks, so none shows $C<2$ on the triangle. The reason is more basic than cuts. The Bell triangle has the same entropy on every group of users as two GHZ states shared by all three users: each user has entropy $2$, and so does each pair. Two such GHZ states give $2$ key bits per copy, so no upper bound on $C$ that uses only the entropies of groups of users, with cut bounds as a special case, can be smaller than $2$ on the triangle. A converse must use information beyond these entropies. We show that $3/2$ is the capacity.

\subsection{Main result}\label{sec:intro-result}

Measuring every GHZ state in the computational basis turns the resource into a classical source of independent uniform bits, each known to the users of its edge, the \emph{measured source}, whose key capacity with public communication is known \cite{CsiszarNarayan04, ChanZheng10}:
\begin{align}
\CZ(\cG)=\min_{\cP:\,|\cP|\ge2}\frac{\sum_{e\in\cE}\big(r_e(\cP)-1\big)}{|\cP|-1}, \label{eq:classical}
\end{align}
where $\cP$ runs over partitions of the users and $r_e(\cP)$ is the number of parts that $e$ meets, that is, the number of holders of $e$ after the users of each part are grouped into one user. We call $\CZ$ the \emph{measured capacity}. The main result is that measuring first is optimal: for every network of independent GHZ states,
\begin{align*}
C=\ED=\CZ,
\end{align*}
under unrestricted LOCC and with vanishing error (Theorem~\ref{thm:main}). The capacity is also achieved with zero error at finite blocklength, as a key rate, as for the measured source \cite{Chan11}, and as a GHZ rate (Section~\ref{sec:achieve}). With vanishing error, and with each GHZ state weighted by the entropy of its label, the capacity result extends to GHZ states with nonuniform labels, and hence to every network of pure bipartite states, each shared by two users, since by the Schmidt decomposition a pure bipartite state is such a GHZ state on two users (Remark~\ref{rem:weights}). Two families serve as running examples. On the \emph{cycle} of $K\ge3$ Bell pairs, of which the Bell triangle is the case $K=3$, the capacity is $K/(K-1)$, so the rate $2$ asked about by Vrana and Christandl is not achievable by LOCC for any cycle. On the \emph{complete network}, with one $J$-user GHZ state on every $J$-subset of users, it is $\binom KJ\frac{J-1}{K-1}$, and $K/2$ for Bell pairs, $J=2$.

For schemes that measure every share before any communication, the partition bound on networks of pure bipartite states was proved by Neumann, Kampermann, Bru{\ss} and Trushechkin \cite[Theorem~4]{Neumann26}. Theorem~\ref{thm:main}, with Remark~\ref{rem:weights}, proves it for all LOCC schemes, in which the users may keep quantum memory and operate on their shares between rounds of public communication. An LOCC converse for products of bipartite pure states was claimed in \cite{TakeokaKRW19}, which was withdrawn because of an unresolved step; we are not aware of a written proof of the unrestricted statement. On the achievability side, the lower bounds of Vrana and Christandl \cite{VranaChristandl19} and of Salek and Winter \cite{SalekWinter22, SalekWinter25} give $\ED\ge\CZ$ with vanishing error.

\subsection{The idea of the converse}\label{sec:intro-idea}

The converse compares two runs of one scheme (Section~\ref{sec:converse}). In the \emph{actual run} the users apply their operations to the GHZ network; in the \emph{product run}, a calculation only, the same local operations are applied to the product of the marginals of the network state, in which the users are independent. The public communication and the keys are the outcome of one measurement whose effects are products of local positive operators. Since every label is held by at least two users, one Cauchy--Schwarz step per label, a case of Finner's inequality \cite{Finner92}, bounds the overlap of the network state with any unit product vector by $1/\sqrt{2^{|\cE|}}$ (Lemma~\ref{lem:twoholder}). So every outcome is at most $2^{\sum_e(|e|-1)}$ times as likely in the actual run as in the product run, where $\sum_e(|e|-1)=\sum_kH(Y_k)-H(Y_1\cdots Y_K)$ is the total correlation of the measured labels, with $Y_k$ the tuple of labels held by user $k$. Averaging the logarithm under the actual run gives an entropy inequality (Lemma~\ref{lem:budget}):
\begin{align*}
\sum_kH(W_k\,|\,F)-H(W_1\cdots W_K\,|\,F)\ \le\ \sum_e(|e|-1),
\end{align*}
where $W_k$ is the key of user $k$ and $F$ the public communication: the total correlation of the keys given the public communication is at most that of the labels. A zero-error $L$-bit key has total correlation $(K-1)L$, so $L\le\sum_e(|e|-1)/(K-1)$, the total-correlation bound of the measured source, now under quantum memory and adaptive communication. Grouping the users of each part of a partition into one user gives \eqref{eq:classical}: edge $e$ then has $r_e(\cP)$ holders, so its weight is $r_e(\cP)-1$, the total correlation of its label among the parts. The proof uses only how the labels are held: how many there are, and that each has at least two holders.

\subsection{Conference key and GHZ distillation}\label{sec:intro-pure}

For every pure source, not only for GHZ networks, conference key and GHZ distillation have the same capacity (Lemma~\ref{prop:equal}): running the key scheme coherently, as in quantum network coding with free classical communication \cite{KobayashiLNR09}, each user measures its residual registers in a basis chosen coherently by its key, so that the outcomes reveal nothing and the residual registers disentangle. In particular, every zero-error key scheme gives exact GHZ distillation at the same rate, which gives the achievability part of Theorem~\ref{thm:main} for GHZ distillation.

\section{Problem Statement and Main Result}\label{sec:model}

We define the problem and state the main result. All logarithms are binary.

\emph{Resources.} For a user subset $e\subseteq[K]$ with $|e|\ge2$, the GHZ state on $e$ is
\begin{align*}
\ket{\GHZ_e}=\frac{1}{\sqrt2}\left(\ket0^{\otimes|e|}+\ket1^{\otimes|e|}\right),
\end{align*}
one qubit per user in $e$; a \emph{Bell pair} is the case $|e|=2$. There are $K\ge2$ users, and a hypergraph $\cG=([K],\cE)$ specifies the resource. Its edge list $\cE$ may contain any user subsets $e$ with $|e|\ge2$, so a network may mix Bell pairs with GHZ states on more users, and may contain the same subset more than once, for several GHZ states on the same users. The resource is
\begin{align}
\ket{\Psi_{\cG}}=\bigotimes_{e\in\cE}\ket{\GHZ_e}, \label{eq:resource}
\end{align}
where different edges occupy distinct registers, also when held by the same user. These are the resources studied in \cite{VranaChristandl17}. The \emph{share} $S_k$ of user $k$ is the set of its registers in the resource. The cycle of Bell pairs has edges $\{k,k+1\}$ with indices modulo $K$; the Bell triangle is the cycle with $K=3$. The complete network $\Psi_{K,J}$ has one GHZ state on every $J$-subset of $[K]$; for $J=2$ it has one Bell pair on every pair of users, and the Bell triangle is $\Psi_{3,2}$.

\emph{Operations.} A scheme with \emph{blocklength} $N$ acts on $N$ copies of a pure $K$-user resource $\ket\psi$, here $\ket{\Psi_\cG}$ and in Lemma~\ref{prop:equal} any pure state, that is, on $\ket\psi^{\otimes N}$, with any finite number $T$ of rounds of classical communication over an authenticated public channel, and with no bound on how $T$ grows with $N$.
\begin{itemize}
\item \emph{Rounds.} In round $t\in[T]$, one user, determined by the messages of rounds $1,\dots,t-1$, applies measurement operators $\{O_x\}_x$ with a finite set of outcomes, chosen as a function of those messages, to its registers and broadcasts the outcome $X_t$. Each $O_x$ is a linear map from the registers of that user to its registers after the round, which may include ancillas, and $\sum_xO_x^\dagger O_x=I$, the identity operator: outcome $x$ on a state $\ket\phi$ has probability $\|O_x\ket\phi\|^2$, with $\|\cdot\|$ the Euclidean norm of a vector, and leaves the unnormalized state $O_x\ket\phi$.\footnote{This is the general local operation with a classical outcome, since a register discarded in a scheme can be kept unused instead. A quantum instrument whose element for outcome $x$ has Kraus operators $O_{x,j}$ becomes the single operator $O_x=\sum_j\ket j_{\rm env}\otimes O_{x,j}$ when the user keeps the environment register, with $O_x^\dagger O_x=\sum_jO_{x,j}^\dagger O_{x,j}$, and discarding the environment gives back the instrument element $\rho\mapsto\Tr_{\rm env}[O_x\rho O_x^\dagger]$.}
\item \emph{Key extraction.} After round $T$, user $k$ applies a measurement with positive operators $\{M_{k,w}\}_w$, $\sum_wM_{k,w}=I$, to its final registers $A_k$, chosen as a function of the \emph{public communication} $F=(X_1,\dots,X_T)$; its outcome is the key $W_k\in\{0,1\}^L$.
\end{itemize}
Public shared randomness is covered by letting one user broadcast it. Given $F=f=(x_1,\dots,x_T)$, the measurement of each round is fixed by the earlier messages; write $O^{(t)}_{x_t}$ for the operator of outcome $x_t$ in round $t$. With $t_1<\dots<t_{T_k}$ the rounds in which user $k$ acts, the operator of user $k$ on the \emph{branch} $f$ is
\begin{align*}
V_{k,f}\ket\phi_{S_k^N}\ =\ O^{(t_{T_k})}_{x_{t_{T_k}}}\cdots O^{(t_2)}_{x_{t_2}}O^{(t_1)}_{x_{t_1}}\ket\phi_{S_k^N},
\end{align*}
a linear map that takes a state $\ket\phi$ of the share $S_k^N$ of user $k$ in the $N$ copies to an unnormalized state of its final registers $A_k$, on which it measures its key. We set $V_{k,f}=I$, with $A_k=S_k^N$, if user $k$ never acts. For a fixed public communication $f$, the operators of different users commute as maps on separate tensor factors, so the operator of the branch $f$ factors as $V_f=\bigotimes_kV_{k,f}$. The branch ends in the unnormalized state $V_f\ket\psi^{\otimes N}$, and the public communication $f$ and the keys $\boldsymbol w=(w_1,\dots,w_K)$ of $W=(W_1,\dots,W_K)$ occur with probabilities
\begin{align*}
P_F(f)=\Tr[V_f\psi^{\otimes N}V_f^\dagger],\qquad P_{FW}(f,\boldsymbol w)=\Tr[\Gamma_{f,\boldsymbol w}\psi^{\otimes N}],
\end{align*}
with $\psi=\ket\psi\!\bra\psi$ the density operator of the source and the local product effects
\begin{align}
\Gamma_{f,\boldsymbol w}\ =\ \bigotimes_k\Gamma_{k,f,w_k},\qquad \Gamma_{k,f,w}\ =\ V_{k,f}^\dagger M_{k,w}V_{k,f}. \label{eq:effects}
\end{align}
The effects sum to $I$: $\sum_wM_{k,w}=I$ reduces the sum to $\sum_fV_f^\dagger V_f$, which is $I$ by $\sum_xO_x^\dagger O_x=I$, summing over the outcome of the last round first, then over that of the round before, and so on. The converse uses only the product form \eqref{eq:effects} of the effects. There is no quantum communication, no pre-shared secret randomness and no free entanglement beyond \eqref{eq:resource}. The eavesdropper observes $F$ and nothing else.

\emph{Requirements.} For $\varepsilon\ge0$ the keys must satisfy
\begin{alignat}{2}
&\text{Correctness:}\qquad &&\Pr\{W_1=\cdots=W_K\}\ge1-\varepsilon, \label{eq:correct}\\
&\text{Secrecy:} &&\tfrac12\|P_{FW_1}-P_F\otimes U_L\|_1\le\varepsilon, \label{eq:secret}
\end{alignat}
where $P_{FW_1}$ is the joint distribution of $(F,W_1)$, a marginal of $P_{FW}$, $P_F$ is the distribution of $F$ given above, $U_L$ is the uniform distribution on $\{0,1\}^L$, and $\|\cdot\|_1$ is the $\ell_1$ norm, so that the left side of \eqref{eq:secret} is the total variation distance; for operators, $\|\cdot\|_1$ is the trace norm, the sum of the singular values, which for Hermitian operators is the sum of the absolute values of the eigenvalues, and $\frac12\|\cdot\|_1$ is the trace distance; the scheme is \emph{zero-error} if $\varepsilon=0$.

\begin{definition}\label{def:cap}
The conference key capacity $C(\psi)$ of a pure $K$-user state $\ket\psi$ is the supremum of the rates $R$ for which there are schemes on $\ket\psi^{\otimes N}$ with $L\ge RN$ and errors $\varepsilon_N\to0$. The GHZ distillation capacity $\ED(\psi)$ is defined in the same way for schemes in which key extraction is replaced by a local operation of each user, chosen as a function of $F$, whose joint output $\sigma_f$ given $F=f$ satisfies $\sum_fP_F(f)\frac12\|\sigma_f-\GHZ_{[K]}^{\otimes L}\|_1\le\varepsilon_N$, with $\ket{\GHZ_{[K]}}$ the GHZ state of all users and $\GHZ_{[K]}$ its density operator.\footnote{The distance is taken given each $f$ and then averaged, since $F$ is public and known to all users.}
\end{definition}

Measuring a distilled GHZ state in the computational basis gives one key bit. Measurement does not increase trace distance, so measuring the output of a GHZ distillation scheme with error $\varepsilon$ gives keys with errors at most $\varepsilon$ in \eqref{eq:correct} and \eqref{eq:secret}, and $\ED(\psi)\le C(\psi)$. Measuring every edge of \eqref{eq:resource} in the computational basis gives the measured source, a classical source of independent uniform bits, the labels of the edges, each known to the holders of its edge. Its key capacity with public communication is the measured capacity $\CZ(\cG)$, given by \eqref{eq:classical} \cite{ChanZheng10}. The \emph{finest partition} is the one into single users. The main result is the following.

\begin{theorem}[Capacity of GHZ networks]\label{thm:main}
For every hypergraph $\cG$,
\begin{align*}
C(\Psi_\cG)=\ED(\Psi_\cG)=\CZ(\cG).
\end{align*}
\end{theorem}

The capacity is also achieved with zero error at finite blocklength, as a key rate \cite{Chan11} and hence, by Lemma~\ref{prop:equal}, as a GHZ rate (Section~\ref{sec:achieve}).

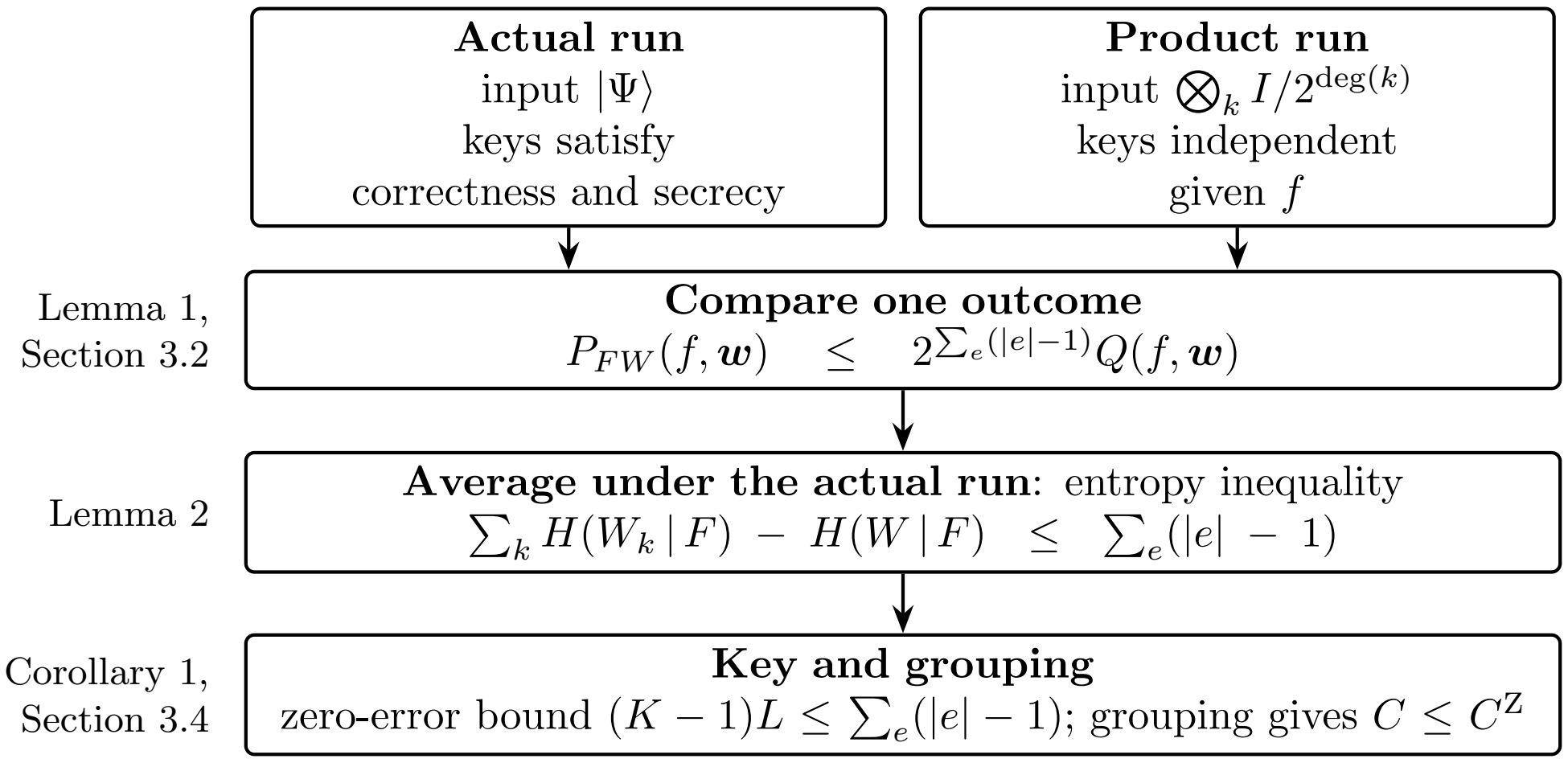
\begin{figure}[t]
\centering
\begin{tikzpicture}[
  box/.style={rectangle, rounded corners=2pt, draw, thick, align=center, font=\small, inner sep=3pt, text width=10.6cm},
  run/.style={rectangle, rounded corners=2pt, draw, thick, align=center, font=\small, inner sep=3pt, text width=5cm},
  lab/.style={font=\footnotesize, align=right, text width=2.2cm, anchor=east, overlay},
  arr/.style={-{Stealth}, thick}]
\node[run] (act) at (-2.75,0) {\textbf{Actual run}\\ input $\ket\Psi$\\ keys satisfy\\ correctness and secrecy};
\node[run] (prod) at (2.75,0) {\textbf{Product run}\\ input $\bigotimes_kI/2^{\deg(k)}$\\ keys independent\\ given $f$};
\node[box] (branch) at (0,-1.75) {\textbf{Compare one outcome}\\ $P_{FW}(f,\boldsymbol w)\le2^{\sum_e(|e|-1)}Q(f,\boldsymbol w)$};
\node[box] (avg) at (0,-3.25) {\textbf{Average under the actual run}: entropy inequality\\ $\sum_kH(W_k\,|\,F)-H(W\,|\,F)\le\sum_e(|e|-1)$};
\node[box] (key) at (0,-4.75) {\textbf{Key and grouping}\\ zero-error bound $(K-1)L\le\sum_e(|e|-1)$; grouping gives $C\le\CZ$};
\draw[arr] (act.south) -- (branch.north -| act.south);
\draw[arr] (prod.south) -- (branch.north -| prod.south);
\draw[arr] (branch) -- (avg);
\draw[arr] (avg) -- (key);
\node[lab] at ($(branch.west)+(-0.15,0)$) {Lemma~\ref{lem:twoholder}, Section~\ref{sec:product}};
\node[lab] at ($(avg.west)+(-0.15,0)$) {Lemma~\ref{lem:budget}};
\node[lab] at ($(key.west)+(-0.15,0)$) {Corollary~\ref{cor:bell}, Section~\ref{sec:simulate}};
\end{tikzpicture}
\caption{The converse. Both runs apply the same measurement, with local product effects, to inputs with the same marginals; the remaining steps turn the comparison of one outcome into an entropy inequality.}
\label{fig:flow}
\end{figure}

The following example proves the capacities of the two running examples stated in Section~\ref{sec:intro-result}, the cycles and the complete networks.

\begin{example}[Cycles and complete networks]\label{ex:families}
On the cycle of $K\ge3$ Bell pairs, a partition $\cP$ is crossed by at least $|\cP|$ edges, so its value is at least $|\cP|/(|\cP|-1)\ge K/(K-1)$, the value of the finest partition; hence the capacity is $K/(K-1)$, and $3/2$ on the triangle. Using each of the $K$ spanning paths of the cycle once on $K-1$ copies, every edge is used $K-1$ times, and fusing each path into a global GHZ state, by broadcasting parities of adjacent labels (computed coherently for GHZ distillation, Section~\ref{sec:achieve}), achieves this rate with zero error. On the complete networks the finest partition is optimal. An edge misses a part of $n$ users exactly when it lies among the other $K-n$ users, which happens for $\binom{K-n}{J}$ edges, so a partition $\cP$ with part sizes $n_1,\dots,n_{|\cP|}$ has value
\begin{align*}
\binom KJ-\frac{1}{|\cP|-1}\sum_i\binom{K-n_i}{J}.
\end{align*}
Since $n\mapsto\binom{K-n}{J}$ is convex, for a fixed number of parts this value is smallest when one part holds all but $|\cP|-1$ users, where it equals $\binom{K-1}{J-1}-\binom{|\cP|-1}{J}/(|\cP|-1)$. The subtracted term equals $\binom{|\cP|-2}{J-1}/J$, which is nondecreasing in $|\cP|$, so the minimum is at $|\cP|=K$, the finest partition, and $C(\Psi_{K,J})=\binom KJ\frac{J-1}{K-1}$, against the minimum cut $\binom{K-1}{J-1}$, the case $|\cP|=2$; for Bell pairs, $J=2$, these are $K/2$ and $K-1$.
\end{example}

\section{Proof of Theorem~\ref{thm:main}: Converse}\label{sec:converse}

The converse compares the actual run of a scheme on the network with its product run, in which the same measurement is applied to the product of the marginals of the network state; Figure~\ref{fig:flow} shows the steps. Section~\ref{sec:twoholder} proves a two-holder inequality for GHZ networks, Section~\ref{sec:product} uses it to compare the probability of each outcome in the two runs, Section~\ref{sec:budget} turns this comparison into an entropy inequality and a key bound, and Section~\ref{sec:simulate} applies the key bound to the network grouped by each partition.

In this section $\ket\Psi=\ket{\Psi_\cG}$ for a GHZ network $\cG=([K],\cE)$, with density operator $\Psi$. User $k$ holds one qubit of each of the GHZ states at $k$; their number $\deg(k)$ is its \emph{degree}, so $\sum_k\deg(k)=\sum_e|e|$, and the density operator of its share $S_k$, the partial trace of $\Psi$ over the other users, is the maximally mixed state $I/2^{\deg(k)}$, since each GHZ state contributes one uniformly distributed label and the GHZ states are independent. Sections~\ref{sec:twoholder}--\ref{sec:budget} consider schemes on one copy of a GHZ network, $N=1$. Since $N$ copies of a GHZ network form a GHZ network in which every edge appears $N$ times, the results also hold for blocklength $N$; Section~\ref{sec:simulate} applies them to $N$ copies of the grouped network.

\subsection{A two-holder inequality}\label{sec:twoholder}

The following inequality bounds the probability that local events at all users occur on a GHZ network; Section~\ref{sec:product} uses it to compare the network with the product of its marginals.

\begin{lemma}[Two holders per label]\label{lem:twoholder}
For positive operators $\Lambda_k$, with $\Lambda_k$ acting on the share $S_k$ of user $k$,
\begin{align}
\bra\Psi\bigotimes_k\Lambda_k\ket\Psi\ \le\ \frac{1}{2^{|\cE|}}\prod_{k=1}^K\Tr\Lambda_k. \label{eq:finner}
\end{align}
\end{lemma}

\begin{proof}
\emph{Step 1: product vectors.} With $\boldsymbol z=(z_e)_e$ the labels of all edges and $\boldsymbol z_{(k)}=(z_e)_{e\ni k}$ those held by user $k$,
\begin{align*}
\ket\Psi\ =\ \frac{1}{\sqrt{2^{|\cE|}}}\sum_{\boldsymbol z}\bigotimes_k\ket{\boldsymbol z_{(k)}}_k.
\end{align*}
For unit vectors $\ket{u_k}$ of $S_k$, let $g_k(\boldsymbol z_{(k)})=|\braket{u_k|\boldsymbol z_{(k)}}|$, so that its $\ell_2$ norm is $\|g_k\|=(\sum_{\boldsymbol z_{(k)}}g_k^2)^{1/2}=\|u_k\|=1$, since the vectors $\ket{\boldsymbol z_{(k)}}$ form the computational basis of $S_k$. Then
\begin{align}
\Big|\Big(\bigotimes_k\bra{u_k}\Big)\ket\Psi\Big|\ &=\ \frac{1}{\sqrt{2^{|\cE|}}}\,\Big|\sum_{\boldsymbol z}\prod_k\braket{u_k|\boldsymbol z_{(k)}}\Big|\ \overset{(a)}{\le}\ \frac{1}{\sqrt{2^{|\cE|}}}\sum_{\boldsymbol z}\prod_kg_k(\boldsymbol z_{(k)})\notag\\
&\overset{(b)}{\le}\ \frac{1}{\sqrt{2^{|\cE|}}}\prod_k\|g_k\|\ =\ \frac{1}{\sqrt{2^{|\cE|}}}, \label{eq:overlap}
\end{align}
where the first equality holds because the inner product of two product vectors is the product of the inner products of their factors, (a) is the triangle inequality $|\sum_{\boldsymbol z}c_{\boldsymbol z}|\le\sum_{\boldsymbol z}|c_{\boldsymbol z}|$ for the terms $c_{\boldsymbol z}=\prod_k\braket{u_k|\boldsymbol z_{(k)}}$, with $|c_{\boldsymbol z}|=\prod_kg_k(\boldsymbol z_{(k)})$, the last equality uses $\|g_k\|=1$, and (b) is the following inequality: for any nonnegative functions $g_k$ of $\boldsymbol z_{(k)}$,
\begin{align}
\sum_{\boldsymbol z}\prod_kg_k(\boldsymbol z_{(k)})\ \le\ \prod_k\|g_k\|. \label{eq:graphholder}
\end{align}
We prove \eqref{eq:graphholder} by induction on the number of edges; it is a case of Finner's inequality \cite{Finner92}. With no edges every $g_k$ is a constant, and \eqref{eq:graphholder} holds with equality. Otherwise pick an edge $e$ with holders $k_1,\dots,k_m$, $m=|e|\ge2$, and write $\boldsymbol z'$ for the labels of all other edges, so that only $g_{k_1},\dots,g_{k_m}$ depend on $z_e$. Let
\begin{align*}
\hat g_{k_i}\ =\ \Big(\sum_{z_e}g_{k_i}^2\Big)^{1/2}\quad(i=1,\dots,m),\qquad \hat g_k\ =\ g_k\quad(k\notin e),
\end{align*}
functions of the labels of user $k$ other than $z_e$. Then
\begin{align*}
\sum_{\boldsymbol z}\prod_kg_k\ &=\ \sum_{\boldsymbol z'}\prod_{k\notin e}g_k\cdot\sum_{z_e}\prod_{i=1}^mg_{k_i}\ \overset{(a)}{\le}\ \sum_{\boldsymbol z'}\prod_{k\notin e}g_k\cdot\prod_{i=3}^m\hat g_{k_i}\cdot\sum_{z_e}g_{k_1}g_{k_2}\\
&\overset{(b)}{\le}\ \sum_{\boldsymbol z'}\prod_{k\notin e}g_k\cdot\prod_{i=1}^m\hat g_{k_i}\ =\ \sum_{\boldsymbol z'}\prod_k\hat g_k\ \overset{(c)}{\le}\ \prod_k\|\hat g_k\|\ \overset{(d)}{=}\ \prod_k\|g_k\|,
\end{align*}
where (a) uses $g_{k_i}\le\hat g_{k_i}$ for $i\ge3$, since one term of a sum of squares is at most the sum; (b) is the Cauchy--Schwarz inequality $\sum_{z_e}g_{k_1}g_{k_2}\le(\sum_{z_e}g_{k_1}^2)^{1/2}(\sum_{z_e}g_{k_2}^2)^{1/2}=\hat g_{k_1}\hat g_{k_2}$; (c) is the induction hypothesis for the network without $e$; and (d) uses $\|\hat g_{k_i}\|=\|g_{k_i}\|$, obtained by summing $\hat g_{k_i}^2$ over the remaining labels held by $k_i$, and $\hat g_k=g_k$ for $k\notin e$.

\emph{Step 2: positive operators.} Write $\Lambda_k=\sum_i\alpha_{k,i}\ket{u_{k,i}}\!\bra{u_{k,i}}$ with eigenvalues $\alpha_{k,i}\ge0$ and orthonormal eigenvectors. Then, by \eqref{eq:overlap} for every choice of eigenvectors,
\begin{align*}
\bra\Psi\bigotimes_k\Lambda_k\ket\Psi\ &=\ \sum_{i_1,\dots,i_K}\prod_k\alpha_{k,i_k}\cdot\Big|\Big(\bigotimes_k\bra{u_{k,i_k}}\Big)\ket\Psi\Big|^2\\
&\le\ \frac{1}{2^{|\cE|}}\sum_{i_1,\dots,i_K}\prod_k\alpha_{k,i_k}\ =\ \frac{1}{2^{|\cE|}}\prod_k\Tr\Lambda_k.\qedhere
\end{align*}
\end{proof}

\subsection{The product run}\label{sec:product}

The product run applies the same scheme to $\bigotimes_kI/2^{\deg(k)}$, the product of the marginals of $\ket\Psi$, in which the users are independent; it is a calculation, not something the users carry out. By \eqref{eq:effects}, it gives $(F,W)$ the distribution
\begin{align}
Q(f,\boldsymbol w)\ =\ \frac{\Tr \Gamma_{f,\boldsymbol w}}{2^{\sum_k\deg(k)}}\ =\ \prod_k\frac{\Tr \Gamma_{k,f,w_k}}{2^{\deg(k)}}. \label{eq:Q}
\end{align}
Given $F=f$ the keys are independent under $Q$: with $Q_F(f)=\prod_k\Tr[V_{k,f}^\dagger V_{k,f}]/2^{\deg(k)}$ and, for $Q_F(f)>0$, $Q_{W_k|F}(w|f)=\Tr \Gamma_{k,f,w}/\Tr[V_{k,f}^\dagger V_{k,f}]$, a distribution over $w$ since $\sum_w\Gamma_{k,f,w}=V_{k,f}^\dagger V_{k,f}$,
\begin{align*}
Q(f,\boldsymbol w)\ =\ Q_F(f)\prod_kQ_{W_k|F}(w_k|f).
\end{align*}

\emph{Comparing one outcome.} By \eqref{eq:effects} and Lemma~\ref{lem:twoholder} with $\Lambda_k=\Gamma_{k,f,w_k}$,
\begin{align}
P_{FW}(f,\boldsymbol w)\ &=\ \bra\Psi \Gamma_{f,\boldsymbol w}\ket\Psi\ \overset{\eqref{eq:finner}}{\le}\ \frac{1}{2^{|\cE|}}\prod_k\Tr \Gamma_{k,f,w_k}\notag\\
&\overset{\eqref{eq:Q}}{=}\ \frac{2^{\sum_k\deg(k)}}{2^{|\cE|}}\,Q(f,\boldsymbol w)\ =\ 2^{\sum_e(|e|-1)}\,Q(f,\boldsymbol w), \label{eq:branch}
\end{align}
where the last step uses $\sum_k\deg(k)-|\cE|=\sum_e|e|-|\cE|=\sum_e(|e|-1)$, since each edge $e$ contributes one qubit to each of its $|e|$ holders.

\subsection{An entropy inequality for GHZ networks}\label{sec:budget}

The comparison \eqref{eq:branch} bounds the \emph{conditional total correlation} $\sum_kH(W_k\,|\,F)-H(W\,|\,F)$ of the keys given the public communication, which equals $(K-1)L$ for a zero-error $L$-bit key.\footnote{If the users measure their labels first, public communication does not increase the total correlation of what they observe \cite{CsiszarNarayan04}, a monotonicity also behind the communication converse for interactive secure summation \cite{SunInteractiveSum26}. Lemma~\ref{lem:budget} gives the same bound for users who keep quantum memory.}

\begin{lemma}[Entropy inequality]\label{lem:budget}
For every scheme on a GHZ network,
\begin{align}
\sum_{k=1}^KH(W_k\,|\,F)-H(W\,|\,F)\ \le\ \sum_{e\in\cE}(|e|-1). \label{eq:budget}
\end{align}
\end{lemma}

\begin{proof}
By \eqref{eq:branch}, $P_{FW}(f,\boldsymbol w)>0$ implies $Q(f,\boldsymbol w)>0$, and the relative entropy satisfies
\begin{align*}
D(P_{FW}\|Q)\ =\ \sum_{f,\boldsymbol w}P_{FW}(f,\boldsymbol w)\log\frac{P_{FW}(f,\boldsymbol w)}{Q(f,\boldsymbol w)}\ \overset{\eqref{eq:branch}}{\le}\ \sum_e(|e|-1).
\end{align*}
Write $P_{W|f}$ and $P_{W_k|f}$ for the distributions of $W$ and $W_k$ given $F=f$ in the actual run, and $Q_{W_k|f}=Q_{W_k|F}(\cdot|f)$. Then
\begin{align*}
D(P_{FW}\|Q)\ &\overset{(a)}{=}\ D(P_F\|Q_F)+\sum_fP_F(f)\,D\Big(P_{W|f}\,\Big\|\,\prod_kQ_{W_k|f}\Big)\\
&\overset{(b)}{=}\ D(P_F\|Q_F)+\sum_fP_F(f)\Big[D\Big(P_{W|f}\,\Big\|\,\prod_kP_{W_k|f}\Big)+\sum_kD\big(P_{W_k|f}\,\big\|\,Q_{W_k|f}\big)\Big]\\
&\overset{(c)}{\ge}\ \sum_fP_F(f)\,D\Big(P_{W|f}\,\Big\|\,\prod_kP_{W_k|f}\Big)\ \overset{(d)}{=}\ \sum_kH(W_k\,|\,F)-H(W\,|\,F),
\end{align*}
where (a) is the chain rule for relative entropy with $Q_{W|F}(\cdot|f)=\prod_kQ_{W_k|f}$; (b) follows by taking the expectation under $P_{W|f}$ of $\log\frac{P_{W|f}}{\prod_kQ_{W_k|f}}=\log\frac{P_{W|f}}{\prod_kP_{W_k|f}}+\sum_k\log\frac{P_{W_k|f}}{Q_{W_k|f}}$; (c) drops relative entropies, which are nonnegative; and (d) uses $D(P_{W|f}\|\prod_kP_{W_k|f})=\sum_kH(W_k\,|\,F=f)-H(W\,|\,F=f)$ and averages over $f$.
\end{proof}

Lemma~\ref{lem:budget} bounds the key length as follows.

\begin{corollary}[Key bound on GHZ networks]\label{cor:bell}
A scheme on a GHZ network that gives an $L$-bit key, $L\ge1$, with errors $0\le\varepsilon\le1/2$ in \eqref{eq:correct}, \eqref{eq:secret} satisfies
\begin{align*}
(K-1)L\big(1-2[\varepsilon+h_2(\varepsilon)]\big)\ \le\ \sum_{e\in\cE}(|e|-1),
\end{align*}
with $h_2$ the binary entropy. In particular $(K-1)L\le\sum_e(|e|-1)$ for a zero-error scheme.
\end{corollary}
\begin{proof}
By the chain rule, with $W=(W_1,\dots,W_K)$,
\begin{align*}
\sum_{k=1}^KH(W_k\,|\,F)-H(W\,|\,F)\ &=\ \sum_{k=2}^KI(W_k;W_1\cdots W_{k-1}\,|\,F)\ \ge\ \sum_{k=2}^KI(W_k;W_1\,|\,F)\\*
&=\ \sum_{k=2}^K\big[H(W_1\,|\,F)-H(W_1\,|\,W_k,F)\big].
\end{align*}
Secrecy \eqref{eq:secret} with continuity of entropy \cite[Theorem~10.7.4]{Wilde17}, applied for each $f$ and averaged over $f$ using the concavity of $h_2$, gives $H(W_1\,|\,F)\ge(1-\varepsilon)L-h_2(\varepsilon)$, and correctness \eqref{eq:correct} with Fano's inequality \cite[Theorem~2.10.1]{CoverThomas06} gives $H(W_1\,|\,W_k,F)\le\varepsilon L+h_2(\varepsilon)$ for every $k\ne1$. So $H(W_1\,|\,F)-H(W_1\,|\,W_k,F)\ge(1-2\varepsilon)L-2h_2(\varepsilon)\ge L(1-2[\varepsilon+h_2(\varepsilon)])$ for every $k\ne1$, since $L\ge1$, and \eqref{eq:budget} gives the claim.
\end{proof}

The following example compares two resources that no bound based on the entropies of groups of users can distinguish.

\begin{example}[Bell triangle and two GHZ states]\label{ex:triangle}
Consider zero-error keys for $K=3$ users, where a common key bit costs $K-1=2$ units of total correlation. The Bell triangle, the cycle of Example~\ref{ex:families} with $K=3$, has three edges $e=\{1,2\},\{1,3\},\{2,3\}$, each a Bell pair held by $|e|=2$ users, so its total correlation per copy is $\sum_e(|e|-1)=3\cdot1=3$, and Corollary~\ref{cor:bell} gives $2L\le3N$, that is, $L\le\frac32N$. Two GHZ states shared by the three users, two copies of $\Psi_{3,3}$, have two parallel edges, both on $\{1,2,3\}$, each held by $|e|=3$ users, so $\sum_e(|e|-1)=2\cdot2=4$, and Corollary~\ref{cor:bell} gives $2L\le4N$, that is, $L\le2N$. Both rates are achieved with zero error, although the two resources have the same entropy on every group of users.
\end{example}

\subsection{Grouping by a partition}\label{sec:simulate}

\begin{proof}[Proof of the converse in Theorem~\ref{thm:main}]
Since $\ED\le C$, it suffices to bound $C$. Fix a partition $\cP$ with $|\cP|\ge2$ parts, and group the users of each part into one user, with the part containing user $1$ using $W_1$; this only enlarges the set of schemes and keeps correctness and secrecy. A GHZ state inside one part can be omitted, since the part can prepare it, and, up to a local unitary, the qubits of a GHZ state held by one part act as one qubit. For a scheme on $N$ copies of $\Psi_\cG$ with $L$ key bits and errors $\varepsilon_N$, the $N$ copies then form a GHZ network on $|\cP|$ users with $N$ copies of each remaining edge $e$, now with $r_e(\cP)$ holders, and Corollary~\ref{cor:bell} with $|\cP|$ in place of $K$ gives
\begin{align*}
(|\cP|-1)L\big(1-2[\varepsilon_N+h_2(\varepsilon_N)]\big)\ \le\ N\sum_e\big(r_e(\cP)-1\big).
\end{align*}
As $\varepsilon_N\to0$ the factor $1-2[\varepsilon_N+h_2(\varepsilon_N)]$ tends to $1$, so no a priori bound on $L/N$ is needed. Dividing by $N$, letting $N\to\infty$, and minimizing over $\cP$ gives $C(\Psi_\cG)\le\CZ(\cG)$. For a zero-error scheme the factor is $1$, so $L/N\le\CZ(\cG)$ at every blocklength $N$.
\end{proof}

\begin{remark}[Nonuniform labels]\label{rem:weights}
Theorem~\ref{thm:main} extends, with vanishing error, to GHZ states $\sum_z\sqrt{p_e(z)}\bigotimes_{k\in e}\ket{u_{e,k}(z)}$ with any finite label distribution $p_e$ and, for each holder $k$, orthonormal vectors $\{\ket{u_{e,k}(z)}\}_z$; by the Schmidt decomposition, these include every pure bipartite state, with $p_e$ the squared Schmidt coefficients. Then \eqref{eq:classical} holds with each edge weighted by $H(p_e)$. Measuring each edge in its label basis gives independent labels with entropies $H(p_e)$, so the classical partition formula \cite{CsiszarNarayan04, ChanZheng10} achieves the weighted value of \eqref{eq:classical} as a key rate, and Lemma~\ref{prop:equal} gives the same lower bound for GHZ distillation. For the converse, entanglement dilution \cite[Section~IV.A]{Bennett00} prepares each such state by LOCC from standard GHZ states on the same holders at rate $H(p_e)$, with vanishing error. Dilution from a GHZ network with $m_e$ copies of each edge, $m_e/N\to H(p_e)$, followed by any GHZ distillation scheme for the nonuniform network gives a scheme on that GHZ network with vanishing error; measuring its output gives keys, and the finite-error bound of Section~\ref{sec:simulate}, with $N$ copies of each edge replaced by $m_e$ copies, which vary with $N$, gives the weighted value of \eqref{eq:classical} after dividing by $N$. By Lemma~\ref{prop:equal}, the same upper bound holds for conference key.
\end{remark}

\section{Proof of Theorem~\ref{thm:main}: Achievability}\label{sec:achieve}

Measuring every GHZ state in the computational basis and running an optimal classical key agreement scheme on the measured source achieves $\CZ(\cG)$ as a key rate \cite[Theorem~1 and Proposition~1]{CsiszarNarayan04}, \cite{ChanZheng10}, so $C(\Psi_\cG)\ge\CZ(\cG)$. Linear communication for omniscience achieves it with zero error at finite blocklength for finite linear sources, which include the measured source \cite{Chan11}. GHZ distillation follows from key agreement by the following result, which holds for every finite-dimensional pure source $\ket\psi$ of $K$ users, with the model of Section~\ref{sec:model}.

\begin{lemma}[From key to GHZ distillation]\label{prop:equal}
For every finite-dimensional pure source in the model of Section~\ref{sec:model}, $\ED(\psi)=C(\psi)$. A zero-error $L$-bit key agreement scheme can be extended by LOCC to give $\ket{\GHZ_{[K]}}^{\otimes L}$ exactly, and a key agreement scheme with errors $\varepsilon$ in \eqref{eq:correct} and \eqref{eq:secret} to give outputs $\sigma_{f'}$ with $\sum_{f'}P_{F'}(f')\frac12\|\sigma_{f'}-\GHZ_{[K]}^{\otimes L}\|_1\le2\sqrt{\varepsilon}$, where $F'=(X_1,\dots,X_{T+K})$ is the public communication $F$ of the key agreement scheme followed by the outcomes $X_{T+1},\dots,X_{T+K}$ broadcast in the extension; this error tends to $0$ as $\varepsilon\to0$.
\end{lemma}

The idea is the following; the proof is in Appendix~\ref{app:equal}. Let the users run the key agreement scheme coherently, keeping all registers. Given the public communication, a zero-error key leaves the state $\frac{1}{\sqrt{2^L}}\sum_w\ket w^{\otimes K}\ket{\chi_w}$, where the residual state $\ket{\chi_w}$ of all other registers may be entangled and may depend on the key value $w$. One user at a time measures its residual registers in a basis that it chooses coherently, as a function of its own key register: a basis that overlaps every eigenvector of its residual state for that key value equally, obtained from the eigenvectors by the discrete Fourier transform. Every outcome then has the same probability for every key value, so the broadcast outcomes reveal nothing about the key, the superposition over $w$ stays uniform, and after the last user only a known phase for each $w$ remains, which one user removes.

Applied to the zero-error scheme of \cite{Chan11}, Lemma~\ref{prop:equal} gives $\ED(\Psi_\cG)\ge\CZ(\cG)$, also with zero error at finite blocklength. Together with the converse of Section~\ref{sec:converse}, this completes the proof of Theorem~\ref{thm:main} and of the zero-error statement after it.

For a linear scheme in which every user recovers all labels and the linear forms defining the key are a basis of a complement of the row space of the broadcast forms, so that the public communication $f$ and the key $w$ together determine all labels, as in \cite{Chan11} and Example~\ref{ex:lift}, the extension is simple. Instead of measuring its labels, each user computes what it would broadcast into an ancilla, measures only the ancilla and broadcasts the outcome. At fixed $f$, the labels of each user are an affine function of $w$ from which the user recovers $w$, so an invertible affine map of the register of that user takes its labels to $(w,0)$; this map permutes the basis states and is therefore unitary. The residual registers are then $\ket0$, so the users need not measure them in the Fourier basis and correct a phase, as in the proof of Lemma~\ref{prop:equal}. The following example shows this for the Bell triangle.

\begin{example}[Two copies of the Bell triangle]\label{ex:lift}
For the cycle of Example~\ref{ex:families} with $K=3$, write the labels of the two copies as $\boldsymbol z=(a_1,a_2,b_1,b_2,c_1,c_2)\in\F_2^6$, with $a_i$, $b_i$, $c_i$ the labels of the Bell pairs on $\{1,2\}$, $\{1,3\}$, $\{2,3\}$ in copy $i$. User $1$ holds $(a_1,a_2,b_1,b_2)$, user $2$ holds $(a_1,a_2,c_1,c_2)$ and user $3$ holds $(b_1,b_2,c_1,c_2)$, in this order in their registers. The two copies are
\begin{align*}
\ket{\Psi}^{\otimes2}=\frac{1}{\sqrt{2^6}}\sum_{\boldsymbol z\in\F_2^6}\ket{a_1a_2b_1b_2}_1\ket{a_1a_2c_1c_2}_2\ket{b_1b_2c_1c_2}_3.
\end{align*}

\emph{Communication.} As in Figure~\ref{fig:triangle}(b), user $k$ broadcasts $x_k$ in round $k$, with
\begin{align*}
f=(x_1,x_2,x_3)=(a_1+b_1,\ a_2+c_1,\ b_2+c_2)=\mathbf H\boldsymbol z,\qquad \mathbf H=\begin{pmatrix}1&0&1&0&0&0\\0&1&0&0&1&0\\0&0&0&1&0&1\end{pmatrix},
\end{align*}
with the columns of $\mathbf H$ ordered as in $\boldsymbol z$.

\emph{Key.} The key is $w=(a_1,a_2,b_2)$. Given $f$, the other labels are
\begin{align}
b_1=a_1+x_1,\qquad c_1=a_2+x_2,\qquad c_2=b_2+x_3, \label{eq:triparam}
\end{align}
so all six labels are determined by $w$ and $f$. The broadcasts $x_1,x_2,x_3$ contain $b_1,c_1,c_2$ respectively, which are not in $w$. Hence $a_1,a_2,b_2,x_1,x_2,x_3$ are linearly independent functions of the labels, $(w,f)$ is uniform on $\F_2^6$, and the key is uniform and independent of $f$: $C\ge3/2$.

\emph{Recovery.} Each user recovers the entries of $w$ it lacks from \eqref{eq:triparam}: user $1$ holds $w$, user $2$ holds $a_1,a_2$ and recovers $b_2=c_2+x_3$, and user $3$ holds $b_2$ and recovers $a_1=b_1+x_1$ and $a_2=c_1+x_2$.

\emph{Ancillas.} The quantum scheme carries out the same steps on the superposition, without measuring the labels. Each user appends an ancilla qubit in $\ket0$ and adds its broadcast to it; for user $1$, $\ket{a_1a_2b_1b_2}_1\ket{0}\mapsto\ket{a_1a_2b_1b_2}_1\ket{a_1+b_1}$, and similarly $a_2+c_1$ for user $2$ and $b_2+c_2$ for user $3$. This is realized by a unitary, the invertible map of the basis states that adds $a_1+b_1$ to the ancilla value. The labels are unchanged, and the three ancillas hold $\mathbf H\boldsymbol z$ in each term:
\begin{align*}
&\frac{1}{\sqrt{2^6}}\sum_{\boldsymbol z\in\F_2^6}\ket{a_1a_2b_1b_2}_1\ket{a_1a_2c_1c_2}_2\ket{b_1b_2c_1c_2}_3\ket{\mathbf H\boldsymbol z}_{\rm anc}\\
&=\sum_{f\in\F_2^3}\frac{1}{\sqrt{2^6}}\sum_{\boldsymbol z:\,\mathbf H\boldsymbol z=f}\ket{a_1a_2b_1b_2}_1\ket{a_1a_2c_1c_2}_2\ket{b_1b_2c_1c_2}_3\ket{f}_{\rm anc}.
\end{align*}
By \eqref{eq:triparam}, the vectors $\boldsymbol z$ with $\mathbf H\boldsymbol z=f$ are given by the $2^3$ values of $w$. The users measure only the ancillas and broadcast the outcome $f$, so $f$ becomes public as before, with probability $2^3/2^6=2^{-3}$ for every $f$. The labels are not measured: all terms with $\mathbf H\boldsymbol z=f$ had the same amplitude $\frac{1}{\sqrt{2^6}}$, so after renormalizing they have amplitude $\frac{1}{\sqrt{2^3}}$, and with $b_1,c_1,c_2$ written through \eqref{eq:triparam} the state is
\begin{align*}
\frac{1}{\sqrt{2^3}}\sum_{a_1,a_2,b_2}\ket{a_1,a_2,a_1{+}x_1,b_2}_1\ket{a_1,a_2,a_2{+}x_2,b_2{+}x_3}_2\ket{a_1{+}x_1,b_2,a_2{+}x_2,b_2{+}x_3}_3.
\end{align*}

\emph{Extracting the key.} No user knows $w$, but every user knows $f$, and hence how its register depends on $w$ in every term. In the key agreement, user $k$ computes $w$ from its labels and $f$; here it applies the same computation inside the ket, by an invertible map of the four bits $(y_1,y_2,y_3,y_4)$ of its register that depends on $f$ and sends its register in every term to $(w,0)$:
\begin{itemize}
\item user $1$: $(y_1,y_2,y_3,y_4)\mapsto(y_1,\,y_2,\,y_4,\,y_3+y_1+x_1)$;
\item user $2$: $(y_1,y_2,y_3,y_4)\mapsto(y_1,\,y_2,\,y_4+x_3,\,y_3+y_2+x_2)$;
\item user $3$: $(y_1,y_2,y_3,y_4)\mapsto(y_1+x_1,\,y_3+x_2,\,y_2,\,y_4+y_2+x_3)$.
\end{itemize}
An invertible map of the labels inside the ket permutes the computational basis states, so it is unitary. Applied to every term, it leaves
\begin{align*}
\frac{1}{\sqrt{2^3}}\sum_{w\in\F_2^3}\ket{w}_1\ket{w}_2\ket{w}_3\otimes\ket0_1\ket0_2\ket0_3=\ket{\GHZ_{123}}^{\otimes3}\otimes\ket0_1\ket0_2\ket0_3
\end{align*}
for every $f$. Thus $\ED\ge3/2$ on the Bell triangle with zero error, from two copies.
\end{example}

\section{Conclusion}\label{sec:discussion}

On every network of independent GHZ states, under unrestricted LOCC with quantum memory and any number of rounds, measuring every GHZ state first achieves the conference key capacity, and a coherent implementation achieves the same rate for GHZ distillation; the capacity is the measured capacity \eqref{eq:classical}. For every pure source, conference key and GHZ distillation have the same capacity.

These results assume that public classical communication is free, in amount and in number of rounds, that the GHZ states are pure and independent, that the eavesdropper observes only the public communication, and that the users have unlimited local quantum memory and ancillas.

\appendix

\section{Proof of Lemma~\ref{prop:equal}}\label{app:equal}

Measuring the output of a GHZ distillation scheme in the computational basis gives keys with at most the same error, so $\ED\le C$, and it remains to show $C\le \ED$.

\emph{Coherent run.} By Section~\ref{sec:model}, each round of the key agreement scheme is the isometry $\ket\phi\mapsto\sum_xO_x\ket\phi\ket x$ followed by a measurement of the outcome register $\ket x$, so given the public communication $f$, with $P_F(f)>0$, the state of the users is the pure state $V_f\ket\psi^{\otimes N}$, normalized. In place of its key measurement $\{M_{k,w}\}$, user $k$ applies the isometry $\ket\phi_{A_k}\mapsto\sum_wM_{k,w}^{1/2}\ket\phi_{A_k}\ket w_{W_k}$ into a key register $W_k$; measuring $W_k$ in the computational basis would give the same keys, with the same joint distribution with $F$. The state stays pure. So given $f$, user $k$ holds its key register $W_k$ and its final registers $A_k$, the \emph{residual register}, of dimension $d_k$, and the global state is a pure state
\begin{align}
\ket{\psi_f}=\sum_{\boldsymbol w}\sqrt{P_{W|F}(\boldsymbol w|f)}\,\ket{w_1}_{W_1}\cdots\ket{w_K}_{W_K}\ket{\chi_{\boldsymbol w}}_{A_1\cdots A_K}, \label{eq:psif}
\end{align}
with $P_{W|F}(\cdot|f)$ the distribution of the keys given $f$ and normalized states $\ket{\chi_{\boldsymbol w}}$, which absorb the phases.

\emph{Removing the residual registers.} For a zero-error key, only key tuples in which all users hold the same value $w$ occur in \eqref{eq:psif}, each with probability $2^{-L}$, so
\begin{align*}
\ket{\psi_f}=\frac1{\sqrt{2^L}}\sum_w\ket w^{\otimes K}\ket{\chi_w^{(1)}},
\end{align*}
with $\ket{\chi_w^{(1)}}=\ket{\chi_{(w,\dots,w)}}$ the residual state in \eqref{eq:psif} of the key tuple in which all users hold $w$. The users act in the order $k=1,\dots,K$. Before step $k$ the state is $\frac{1}{\sqrt{2^L}}\sum_w\ket w^{\otimes K}\ket{\chi_w^{(k)}}$ with $\ket{\chi_w^{(k)}}$ on $A_k\cdots A_K$. Let $\rho_w$ be the marginal of $\ket{\chi_w^{(k)}}$ on $A_k$, with eigenvectors $\ket{v_{w,j}}$, $j=0,\dots,d_k-1$, and let $G_w$ be the Fourier transform from this eigenbasis to the computational basis; $\rho_w$, its eigenvectors and $G_w$ also depend on the step $k$ and on the outcomes broadcast so far, which the notation suppresses,
\begin{align*}
G_w\ket{v_{w,j}}=\frac{1}{\sqrt{d_k}}\sum_{x=0}^{d_k-1}\omega^{-jx}\ket x,\qquad \omega=e^{2\pi i/d_k}.
\end{align*}
Step $k$ is the following.
\begin{itemize}
\item \emph{Fourier transform.} Controlled by its key register, user $k$ applies $G_w$ to $A_k$, the unitary $\sum_w\ket w\!\bra w_{W_k}\otimes G_w$.
\item \emph{Measurement.} User $k$ measures $A_k$ in the computational basis and broadcasts the outcome $x$ as $X_{T+k}$. Outcome $x$ has probability $\bra xG_w\rho_wG_w^\dagger\ket x=1/d_k$ for every $w$, since $\rho_w$ is diagonal in the eigenbasis and $|\bra xG_w\ket{v_{w,j}}|^2=1/d_k$ for all $j$, so the outcome is independent of the key, and the state becomes $\frac{1}{\sqrt{2^L}}\sum_w\ket w^{\otimes K}\ket{\chi_w^{(k+1)}}$ with the normalized state $\ket{\chi_w^{(k+1)}}=\sqrt{d_k}\,(\bra xG_w\otimes I)\ket{\chi_w^{(k)}}$ of $A_{k+1}\cdots A_K$.
\end{itemize}
After step $K$ the state is $\frac{1}{\sqrt{2^L}}\sum_w\gamma_w\ket w^{\otimes K}$ with $|\gamma_w|=1$, where the phase $\gamma_w$ is determined by $w$, $f$ and the broadcast outcomes. User $1$ removes this phase by applying $\ket w\mapsto\gamma_w^{-1}\ket w$, a diagonal unitary since $|\gamma_w|=1$, which leaves $\ket{\GHZ_{[K]}}^{\otimes L}$. The scheme uses only local operations and broadcast messages, and the broadcast messages are independent of the key.

\emph{Keys with error.} This case follows by continuity. Let $p_f(w)$ be the probability given $f$ that all users hold the key value $w$. Compare $\ket{\psi_f}$ in \eqref{eq:psif} with $\ket{\psi^\circ_f}=\frac{1}{\sqrt{2^L}}\sum_w\ket w^{\otimes K}\ket{\chi_w^{(1)}}$, where $\ket{\chi_w^{(1)}}$ is the residual state in \eqref{eq:psif} of the key tuple in which all users hold $w$, and any normalized state when $p_f(w)=0$, a choice that does not affect the overlap below. This is the state of a zero-error key in the previous step, so the operations constructed there turn it into $\ket{\GHZ_{[K]}}^{\otimes L}$. Only the key tuples in which all users agree contribute to the overlap, with the same residual state in both, so the overlap is $s_f=\braket{\psi^\circ_f|\psi_f}=\sum_w\sqrt{2^{-L}p_f(w)}$. Since $\sqrt{ab}\ge\min\{a,b\}$ and $\sum_w2^{-L}=1$, $1-s_f\le\sum_w(2^{-L}-p_f(w))_+$, with $(a)_+=\max\{a,0\}$, and this is at most the total variation distance of the distribution of $W_1$ given $f$ from uniform plus the probability that the keys disagree given $f$. Averaged over $f$, $\sum_fP_F(f)(1-s_f)\le2\varepsilon$ by \eqref{eq:correct} and \eqref{eq:secret}.

Apply to $\ket{\psi_f}$ the operations constructed for $\ket{\psi^\circ_f}$, including the phase correction, and let $q(f'|f)$ and $\sigma_{f'}$ be the probability and the output for each public communication $f'$ of the extended scheme that extends $f$. With the outcomes of the extension discarded, these operations form a quantum channel $\mathcal N_f$, which maps $\ket{\psi^\circ_f}$ to $\GHZ_{[K]}^{\otimes L}$, since every outcome does, and $\ket{\psi_f}$ to $\sum_{f'}q(f'|f)\sigma_{f'}$. Fidelity does not decrease under quantum channels, so
\begin{align*}
\sum_{f'}q(f'|f)\,\bra{\GHZ_{[K]}}^{\otimes L}\sigma_{f'}\ket{\GHZ_{[K]}}^{\otimes L}\ \ge\ |\braket{\psi^\circ_f|\psi_f}|^2\ =\ s_f^2.
\end{align*}
Since $\frac12\|\sigma-\GHZ_{[K]}^{\otimes L}\|_1\le\sqrt{1-\bra{\GHZ_{[K]}}^{\otimes L}\sigma\ket{\GHZ_{[K]}}^{\otimes L}}$ and the square root is concave, the average of $\frac12\|\sigma_{f'}-\GHZ_{[K]}^{\otimes L}\|_1$ over $f'$ given $f$ is at most $\sqrt{1-s_f^2}\le\sqrt{2(1-s_f)}$, and averaging over $f$, which gives the average over $F'$ in Lemma~\ref{prop:equal}, gives at most
\begin{align*}
\sum_fP_F(f)\sqrt{2(1-s_f)}\ \le\ \sqrt{2\sum_fP_F(f)(1-s_f)}\ \le\ 2\sqrt\varepsilon.
\end{align*}
For a sequence of key agreement schemes with errors $\varepsilon_N\to0$, the extended schemes have errors at most $2\sqrt{\varepsilon_N}\to0$, at the same rate. Hence every key rate is a GHZ rate, and $C\le \ED$.

\bibliography{refs}

\end{document}